\documentstyle[aps]{revtex}
\begin{document}
\title{OPERATOR FOR DESCRIBING POLARIZATION STATES OF A PHOTON\thanks{%
Work supported in part by National Natural Science Foundation of China}}
\author{Hong-Yi Fan$^{a,b,c,d}$ and Qiu-Yu Liu$^d$}
\address{$^a$ CCAST(World Laboratory), P.O.Box 8730, 100080,
Beijing,China\\
$^b$Department of Applied Physics ,Shanghai Jiao Tong University ,Shanghai
200030,China\\
$^c$Department of Material Science and Engineering, University of Science
and Technology of China, Hefei, Anhui 230026, China.\\
$^d$International Center for Theoretical Physics, Strada Costiera
11,34100,
Trieste, Italy.}
\maketitle

\begin{abstract}
Based on the quantized electromagnetic field described by the
Riemann-Silberstein complex vector $F$, we construct the eigenvector set
of $%
F$, which makes up an orthonormal and complete representation. In terms of
$%
F $ we then introduce a new operator which can describe the relative ratio
of the left-handed and right-handed polarization states of a polarized
photon .In $F^{\prime }s$\ eigenvector basis the operator manifestly
exhibits a behaviour which is similar to a phase difference between two
orientations of polarization of a light beam in classical optics.

\end{abstract}

\section{Introduction}

In recent years, single photon states has attracted much attention of
physicist because they exhibit high powerlaw falloff of the energy density
and of the photon detection rates [1-3].The photon states are conveniently
described in terms of the Riemann-Silberstein complex vector of
electromagnetic field [4]. The spatial localization properties of one
photon
states and the decay in time has been analyzed in reference [3]. In the
present letter we seek a new quantum mechanical formalism to determine the
nature of the polarization of a single photon, which is based on the
quantized Riemann-Silberstein complex vector .

As for the polarization of single photon states let us quote the idea of
Dirac [5]: ''When we make the photon meet a tourmaline crystal, we are
subjecting it to an observation. We are observing whether it is polarized
parallel or perpendicular to the optic axis. The effect of making this
observation is to force the photon entirely into the state of parallel or
entirely into the state of perpendicular polarization. It has to make a
sudden jump from being partly in each of these two states to being
entirely
in one or other of them. Which of the two states it will jump into cannot
be
predicted, but is governed only by probability laws. If it jumps into the
parallel state it gets absorbed and if it jumps into the perpendicular
state
it passes through the crystal and appears on the other side preserving
this
state of polarization.''. In this work we want to introduce a new operator
which can describe the relative ratio of the two polarizations
(left-handed
and right-handed) for a photon polarized obliquely to the optic axis.
Moreover,when we establish a new representation $|\xi >_k$for the
quantized
Riemann-Silberstein complex vector of electromagnetic field,it turns out
that in $|\xi >_k$ representation the new operator can manifestly exhibit
a
behaviour which is similar to a phase difference between two orientations
of
polarization for a beam in classical optics. Recall that for a beam of
classical electromagnetic plane wave the electric fields $E_x=A_1\cos
(\tau
+\delta _1),$ $E_y=A_2\cos (\tau +\delta _2)$ can be re-written as
(eliminating $\tau \;$between these two equations) 
\begin{equation}
\left( 
%TCIMACRO{\dfrac{E_x}{A_1}}
%BeginExpansion
{\displaystyle {E_x \over A_1}}
%EndExpansion
\right) ^2+\left( 
%TCIMACRO{\dfrac{E_y}{A_2}}
%BeginExpansion
{\displaystyle {E_y \over A_2}}
%EndExpansion
\right) ^2-2%
%TCIMACRO{\dfrac{E_x}{A_1}}
%BeginExpansion
{\displaystyle {E_x \over A_1}}
%EndExpansion
%TCIMACRO{\dfrac{E_y}{A_2}}
%BeginExpansion
{\displaystyle {E_y \over A_2}}
%EndExpansion
\cos \delta =\sin ^2\delta ,\text{ \ \ }\delta =\delta _2-\delta _1,
\label{1}
\end{equation}
where exp(i$\delta )$ is the phase difference as a parameter describing
the
polarization [6].For example,for a right-handed polarized electric wave,
$%
E_y/E_x=e^{i\delta }=e^{-i\pi /2}.\;$The work is arranged as follows: in
Sec. 2 we construct the eigenvector of the quantized Riemann-Silberstein
complex field. In Sec. 3 we introduce the operator for the description of
photon polarization states in terms of the quantized Riemann-Silberstein
complex vector. In Sec.4 We point out that although our new operator has
nothing to do with optical phase in the usual sense which was discussed in
[7-9] by Dirac,Susskind-Glogower and Lynch,we compare and contrast the
operator with Noh, Fougeres and Mandel (NFM) operationally defined phase
operator [10] [11] because they resemble to each other in form .

\section{Eigenvector of quantized Riemann-Silberstein complex vector}

As shown in [1], the most general one-photon (1ph) state can be described
by
two complex functions of the wave vector 
\begin{equation}
|1ph>=\int d^3kf_{+}({\bf k})a^{\dagger }({\bf k})|00>_k~+~\int
d^3kf_{-}(%
{\bf k})b^{\dagger }({\bf k})|00>_k,  \label{2}
\end{equation}
where $f_{\pm }$ are two components of the photon wave function in
momentum
representation which is normalized to one. 
\begin{equation}
\int d^3k|f_{+}({\bf k})|^2~+~\int d^3k|f_{-}({\bf k})|^2~=~1~.  \label{3}
\end{equation}
and $a^{\dagger }({\bf k})$, $b^{\dagger }({\bf k})$ are the creation
operators of photons with the left-handed and right-handed polarization,
satisfying 
\begin{equation}
\lbrack a({\bf k}),a^{\dagger }({\bf k}^{\prime })]~=~\delta ({\bf
k}~-~{\bf %
k}^{\prime })~=~[b({\bf k}),b^{\dagger }({\bf k}^{\prime })]~.  \label{4}
\end{equation}
Eqs. (2-4) coincide with Dirac's idea [5] that :''It is supposed that a
photon polarized obliquely to the optic axis may be regarded as being
partly
in the state of polarization perpendicular to the axis. The state of
oblique
polarization may be considered as the result of some kind of superposition
process applies to the two states of parallel and perpendicular
polarization. This implies a certain special kind of relationship between
the various states of polarization, a relationship similar to that between
polarized beams in classical optics, but which is now to be applied, not
to
beams, {\it but to the states of polarization of one particular photon}.
This relationship allows any state of polarization to be resolved into, or
expressed as a superposition of, any two mutually perpendicular states of
polarization.''

The photon states are conveniently described in terms of the
Riemann-Silberstein complex vector ${\bf F}$ (RS vector) composed of
electric displacement and the magnetic induction vectors [4] , 
\begin{equation}
{\bf F}({\bf r},t)~=~\frac{{\bf D}({\bf r},t)}{\sqrt{2\epsilon
_0}}~+~i\frac{%
{\bf B}({\bf r},t)}{\sqrt{2\mu _0}}.~  \label{5}
\end{equation}
The square roots of $\epsilon $ and $\mu $ are needed to match the
dimensions of the two terms and an additional factor of $\sqrt{2}$ is
introduced to make the modulus of ${\bf F}$ equal simply to the energy
density 
\begin{equation}
H_{CL}({\bf r},t)~=~\frac{{\bf D}^2({\bf r},t)}{2\epsilon _0}~+~\frac{{\bf
B}%
^2({\bf r},t)}{2\mu _0}~=~{\bf F}^{\dagger }({\bf r},t)\cdot {\bf F}({\bf
r}%
,t)~.  \label{6}
\end{equation}
After quantizing the electromagnetic field, the RS vector becomes the
field
operator $\hat {{\bf F}}({\bf r},0)$ (we consider $t=0$ case in
Schr\"odinger picture) 
\begin{equation}
\hat F({\bf r},0)~=~\int d^3k{\ \sqrt{\frac{\hbar c{\bf k}}{(2\pi
)^3}}}e(%
{\bf k})F({\bf k},{\bf r})  \label{7}
\end{equation}
where $e({\bf k})$ is an unit polarization vector, and 
\begin{equation}
F({\bf k},{\bf r})~=~a({\bf k})f_k~+~b^{\dagger }({\bf k})f_k^{*},~~~F^{%
\dagger }({\bf k},{\bf r})~=~a^{\dagger }({\bf k})f_k^{*}~+~b({\bf k})f_k
\label{8}
\end{equation}
here $f_k=e^{i{\bf k}\cdot {\bf r}}$.We now introduce the eigenvectors of
$F(%
{\bf k},{\bf r})$ : 
\begin{equation}
|\xi >_k~=~exp(-\frac{|\xi |^2}2+\xi f_k^{-1}a^{\dagger }({\bf k})+\xi
^{*}f_k^{-1}b^{\dagger }({\bf k})-a^{\dagger }({\bf k})b^{\dagger }({\bf
k}%
)f_k^{-2})|00>_k  \label{9}
\end{equation}
where $\xi =\xi _1+i\xi _2$ is a complex number and the vacuum state
$|00>_k$
is annihilated by both $a({\bf k})$ and $b({\bf k})$. We prove here that
it
is an eigenstate of the operator $F({\bf k},{\bf r})$ by acting $a({\bf
k})$
on $|\xi >_k$. 
\begin{equation}
\begin{array}{l}
a({\bf k})|\xi >_k=~{\ [a({\bf k}),exp(-\frac{|\xi |^2}2+\xi
f_k^{-1}a^{\dagger }({\bf k})+\xi ^{*}f_k^{-1}b^{\dagger }({\bf k}%
)-a^{\dagger }({\bf k})b^{\dagger }({\bf k})f_k^{-2})]}|00>_k \\ 
\qquad \qquad \quad =~(\xi f_k^{-1}-b^{\dagger }({\bf k})f_k^{-2})|\xi
>_k~.
\end{array}
\label{10}
\end{equation}
It then follows 
\begin{equation}
F({\bf k},{\bf r})|\xi >_k\equiv (a({\bf k})f_k~+~b^{\dagger }({\bf k}%
)f_k^{-1})|\xi >_k~=~\xi |\xi >_k  \label{11}
\end{equation}
On the other hand, by acting $b({\bf k})$ on $|\xi >_k$, we have 
\begin{equation}
b({\bf k})|\xi >_k~=~(\xi ^{*}f_k^{-1}-a^{\dagger }({\bf k})f_k^{-2})|\xi
>_k~.  \label{12}
\end{equation}
which yields 
\begin{equation}
F^{\dagger }({\bf k},{\bf r})|\xi >_k\equiv (b({\bf k})f_k~+~a^{\dagger
}(%
{\bf k})f_k^{-1})|\xi >_k~=~\xi ^{*}|\xi >_k  \label{13}
\end{equation}
Thus $|\xi >_k$ is the common eigenvector of $F$ and $F^{\dagger }$, which
agree with the commutator 
\[
\lbrack F({\bf k},{\bf r}),F^{\dagger }({\bf k},{\bf r})]~=~0. 
\]
Using the technique of integration within an ordered product (IWOP) of
operators [12]] and $|00>_k._k<00|=:exp(-a^{\dagger }({\bf k})a({\bf k}%
)-b^{\dagger }({\bf k})b({\bf k})):$, (where $:~:$ denotes normal
ordering,)
we can perform the following integration (In the following for brevity we
write $a({\bf k})$ as $a$, $f_k$ as $f$ and so on.)

\begin{equation}
\begin{array}{c}
{\int \frac{d^2\xi }\pi |\xi >_k._k<\xi |~=~\int \frac{d^2\xi }\pi }%
:exp[-|\xi |^2+ \\ 
\xi (f^{*}a^{\dagger }+fb)+\xi ^{*}(fa+f^{*}b^{\dagger })-a^{\dagger
}b^{\dagger }f^{-2}-abf^2-a^{\dagger }a-b^{\dagger }b]:~=~{\ {\bf {\Huge
1}}.%
}
\end{array}
\label{14}
\end{equation}

This indicates that $|\xi >_k$ make up a completeness relation. Further,
by
examining Eqs. (11)-(13) we see 
\begin{equation}
\begin{array}{c}
_k<\xi ^{\prime }|F({\bf k},{\bf r})|\xi >_k~=~\xi ~_k<\xi ^{\prime }|\xi
>_k~=~\xi ^{\prime }~_k<\xi ^{\prime }|\xi >_k; \\ 
_k<\xi ^{\prime }|F^{\dagger }({\bf k},{\bf r})|\xi >_k~=~\xi ^{*}~_k<\xi
^{\prime }|\xi >_k~=~\xi ^{\prime }{}^{*}~_k<\xi ^{\prime }|\xi >_k,
\end{array}
\label{15}
\end{equation}
thus $(\xi ^{\prime }-\xi )~_k<\xi ^{\prime }|\xi >_k=(\xi ^{\prime
}{}^{*}-\xi ^{*})~_k<\xi ^{\prime }|\xi >_k=0$ which indicates the
orthonormal property 
\begin{equation}
_k<\xi ^{\prime }|\xi >_k~=~\pi \delta (\xi ^{\prime }-\xi )\delta (\xi
^{\prime }{}^{*}-\xi ^{*}).  \label{16}
\end{equation}

\section{Operator for describing photon polarization states}

For describing polarization states of single photon we now introduce a new
operator 
\begin{equation}
{e^{i\hat \Theta }}\equiv \sqrt{\frac F{F^{\dagger }}}=\sqrt{\frac{%
af+b^{\dagger }f^{*}}{a^{\dagger }f^{*}+bf}}.  \label{17}
\end{equation}
This definition is feasible as $[af+b^{\dagger }f^{*},~a^{\dagger
}f^{*}+bf]=0$ and they can reside within the same square root without
ambiguity. Physically, since $a^{\dagger }({\bf k})$, $b^{\dagger }({\bf
k})$
are creation operators of left-handed and right-handed polarization
respectively, ${\ \frac{af+b^{\dagger }f^{*}}{a^{\dagger }f^{*}+bf}}$
represents the relative ratio of the two polarizations. This
phase-difference between two polarizations which is similar to e$^{i\delta
}$
in classical optics (see Eq.(1)) can be seen more clearly in our $|\xi
>_k$
bases. By using Eqs.(11) and (13) we have 
\begin{equation}
e^{i\hat \Theta }={\ \sqrt{\frac F{F^{\dagger }}}~=~\int \frac{d^2\xi }\pi
~{%
\ \sqrt{\frac \xi {\xi ^{*}}}}|\xi >_k._k<\xi |~=~\int \frac{d^2\xi }\pi
~e^{i\theta }|\xi >_k._k<\xi |}  \label{18}
\end{equation}
where $\xi =|\xi |e^{i\theta }$.e$^{i\theta }$ provides an involved
description of the state of polarization of the field..One can immediately
determine from the value of this ratio the nature of the polarization.
Writing $d^2\xi =|\xi |d|\xi |d\theta $, we can express the expectation
value of $e^{i\hat \Theta }$ in a normalized state $|\psi >$ as 
\begin{equation}
<\psi |e^{i\hat \Theta }|\psi >={\ \int \frac{d^2\xi }\pi ~<\psi
|e^{i\theta
}|\xi >_k{}_k<\xi |\psi >}={\ \int_0^{2\pi }d\theta e^{i\theta }\
\int_0^\infty \frac{|\xi |d|\xi |}\pi |<\xi |\psi >|^2}  \label{19}
\end{equation}
Let $P(\theta )={\ \int_0^\infty \frac{|\xi |d|\xi |}\pi |<\xi |\psi
>|^2}$,
we have 
\begin{equation}
<\psi |e^{i\hat \Theta }|\psi >={\ \int_0^{2\pi }d\theta e^{i\theta }}%
P(\theta ).  \label{20}
\end{equation}
According to one of the postulates of the quantum mechanics [13]: ``When
the
physical quantity A, to which orthonormalized eigenvectors $|u_n>$
associated with the eigenvalue $\omega _n$ correspond, the expectation
value
of A in $|\psi >$ is given by $<\psi |A|\psi >=\Sigma _n|C_n|^2\omega _n$,
where $|C_n|^2=|<u_n|\psi >|^2$ is the probability. `` We have reason to
name $P(\theta )$ in Eq. (20) the probability distribution function
describing the degree of circular polarization,since $e^{i\theta }$ is the
eigenvalue of $e^{i\hat \Theta }$. This shows that $|\xi >_k$ spans a
spectral representation of the operator $\sqrt{\frac F{F^{\dagger }}}$%
.Moreover, by introducing the number difference between left-handed and
right-handed polarization operator $Q=a^{\dagger }a-b^{\dagger }b$, we see 
\begin{equation}
\lbrack Q,F]=-F,~~~~~~~~~~[Q,F^{\dagger }]=F^{\dagger }  \label{21}
\end{equation}
Note that $e^{i\hat \Theta }$ is unitary,we can conclude: 
\begin{equation}
\lbrack Q,e^{i\hat \Theta }]=-e^{i\hat \Theta },~~~~~~~~~~[Q,e^{-i\hat
\Theta
}]=e^{-i\hat \Theta }  \label{22}
\end{equation}
We can also define the Hermitian operator of the angle as 
\begin{equation}
\hat \Theta ~=~\frac 1\pi \int d^2\xi |\xi >_k{}_k<\xi |\theta ,
\label{23}
\end{equation}
then as a result of Eq. (21), we have 
\begin{equation}
e^{i\hat \Theta }Qe^{-i\hat \Theta }~=~Q+1~=~Q+[i\hat \Theta ,~Q]+\frac
1{2!}%
[i\hat \Theta ,[i\hat \Theta ,Q]]+\frac 1{3!}[i\hat \Theta ,[i\hat \Theta
,[i%
\hat \Theta ,Q]]]+......  \label{24}
\end{equation}
which implies formally $[Q,\hat \Theta ]=i$.

\section{Comparison with NFM operationally defined phase operator}

It is worth comparing and contrasting Eq.(17) with the NFM operational
phase
operator,which is based on an eight-port homodyne experiment . As
measurement always involve the difference between two phases, and as an
homodyne experiment usually yields the cosine or sine of phase difference
between two quantum states,the state of the input modes are an arbitrary
two-mode state---the signal state in modes 10 and 1,and a coherent state
(a
local oscillator )---the reference, NFM proposed the cosine phase operator
(
in the limit of a strong local oscillator ) as [10-11][14] 
\begin{equation}
C=\frac{\widehat{X}}{\sqrt{\widehat{X}^2{\tt +}\widehat{P}^2}}=\frac{%
\widehat{x}_1+\widehat{x}_{10}}{\sqrt{(\widehat{x}_1+\widehat{x}_{10})^2+(%
\widehat{p}_1-\widehat{p}_{10})^2}}  \label{25}
\end{equation}
where 
\begin{eqnarray}
x_1 &=&\frac 1{\sqrt{2}}({\tt a}+{\tt a}^{\dagger }),\quad x_{10}=\frac
1{%
\sqrt{2}}({\tt b}+{\tt b}^{\dagger }),\quad p_1=\frac 1{\sqrt{2}i}({\tt
a}-%
{\tt a}^{\dagger }),\quad p_{10}=\frac 1{\sqrt{2}i}({\tt b}-{\tt
b}^{\dagger
}).  \nonumber  \label{26} \\
\lbrack {\tt a},{\tt a}^{\dagger }] &=&[{\tt b},{\tt b}^{\dagger }]=1.
\label{26}
\end{eqnarray}
Substituting Eq. (26) into (25) leads to 
\begin{equation}
\begin{array}{c}
C=\frac{\hat X}{\sqrt{\hat X^2+\hat P^2}}=\frac{{\tt a}+{\tt a}^{\dagger
}+%
{\tt b}+{\tt b}^{\dagger }}{2\sqrt{{\tt aa}^{\dagger }+{\tt bb}^{\dagger
}+%
{\tt ab}+{\tt a}^{\dagger }{\tt b}^{\dagger }}}=\frac 12[\frac{{\tt a}%
^{\dagger }+{\tt b}}{\sqrt{({\tt a}^{\dagger }+{\tt b})({\tt b}^{\dagger
}+%
{\tt a})}}+\frac{{\tt a}+{\tt b}^{\dagger }}{\sqrt{({\tt a}^{\dagger
}+{\tt b%
})({\tt b}^{\dagger }+{\tt a})}}] \\ 
=\frac 12[\sqrt{\frac{{\tt a}^{\dagger }+{\tt b}}{{\tt a}+{\tt b}^{\dagger
}}%
}+\sqrt{\frac{{\tt a}+{\tt b}^{\dagger }}{{\tt a}^{\dagger }+{\tt b}}}]=%
\frac 12(e^{-i\alpha }+e^{i\alpha })=\cos \alpha ,
\end{array}
\label{27}
\end{equation}
where $e^{i\alpha }=\sqrt{\frac{{\tt a}+{\tt b}^{\dagger }}{{\tt a}\dagger
+%
{\tt b}}}$ is comparable in form with the operator $\sqrt{\frac
F{F^{+}}}=%
\sqrt{\frac{af+b^{\dagger }f^{*}}{a^{\dagger }f^{*}+bf}}$ for polarization
states of a photon. However,it must emphasized that although our operator
$F$
resembles in form to the NFM phase operator (see Eq.(27) ) ,their physical
meaning are completely different,since a single photon cannot carry phase
information.

In summary,we have discussed the intrinsic relation between the quantized
Riemann-Silberstein electromagnetic complex vector $F$ and the
determination
of the nature of the polarization of a single photon. We have constructed
$F$%
's orthonormal and complete eigenvector set. Based on this, the new
description reflecting the relative ratio between left-handed and
right-handed polarization states of a photon is established quantum
mechanically. Although our formalism seems to be comparable to NFM
operationally defined phase operator,their physical conceptions are by no
means the same,the operator (17) has absolutely nothing to do with optical
phase in the usual sense,that is as the quantity shifted by a
phase-shifter
in a interferometer.

{\bf References}

\begin{itemize}
\item[1.]  I. Bialynicki-Birula, Phys. Rev. Lett. 80 (1998) 5247 and
references therein

\item[2.]  I. Bialynicki-Birula, ACTA Physica Polonica A86 (1994) 97,
Proceedings of the International Conference ``Quantum Optics III'',
Szczyrk,
Poland, 1993.

\item[3.]  C. Adlard, E.R. Pike, and S. Sarkar, Phys. Rev. Lett. 79 (1997)
1583.

\item[4.]  L. Silberstein, Ann. Phys. (Leipzig) 22, 579 (1907); 24, 783
(1907).

\item[5.]  P.A.M. Dirac, The Principle of Quantum Mechanics,
Oxford,3rd,1958

\item[6.]  M.Born and E.Wolf, Principles of Optics,Pergamon
Press,1980,Oxford

\item[7.]  .P.A.M. Dirac, Proc. Roy. Soc. A114 (London 1927) 243.

\item[8.]  L. Susskind and J. Glogower, Physics 1 (1964) 49;

\item[9.]  R. Lynch, Phys. Rep. 256 (1995) 367.

\item[10.]  Noh,Fongeres and L.Mandel,Phys.Rev.Lett 67 (1991)
1426;Phys.Rev.A 45 (1992) 424;Freyberger et al Phys. Lett.A176 (1993) 41 ;

\item[11.]  Englert et al,Phys.Rev.A51 (1995) R2661;Englert et
al,Phys.Rev.A52 (1995)1704.

\item[12.]  Fan Hong-Yi, H.R. Zaidi and J.R. Klauder, Phys. Rev. D35
(1987)
1831;Fan Hong-Yi and Fan Yue, Phys. Rev. A54 (1996) 958;Fan Hongyi and
J.R.Klauder,Phys.Rev.A49 (1994) 704

\item[13.]  Claude Cohen-Tannoudji et al, Quantum Mechanics Vol. 1, John
Wiley \& Sono, 1997

\item[14.]  Freyberger and L.Mandel, Quant.Semiclass.Opt.7 (1995) 187
\end{itemize}

\end{document}